\documentstyle[PASJadd]{PASJ95}
\draft
\begin{document}               
\title{ASCA Observations of a Nearby and Massive \\ Galaxy Cluster, Abell 3627}
\author{
Takayuki {\sc Tamura}, Yasushi {\sc Fukazawa},
Hidehiro {\sc Kaneda}, Kazuo {\sc Makishima}, Makoto {\sc Tashiro}\\
{\it Department of Physics, School of Science, The University of Tokyo, Hongo, Bunkyo-ku, Tokyo 113-0033}\\   
{\it E-mail(TT): ttamura@amalthea.phys.s.u-tokyo.ac.jp}\\
Yasuo {\sc Tanaka}\thanks{Present address:Max-Planck-Institut f\"ur extraterrestrische Physik, D-85740 Garching, Germany},\\
{\it The Institute of Space and Astronautical Science, 3-1-1 Yoshinodai, Sagamihara, Kanagawa 229-8510}\\
and\\
Hans {\sc B\"ohringer}\\
{\it Max-Planck-Institut f\"ur extraterrestrische Physik, D-85740 Garching, Germany}\\   
}

\abst{
The results obtained from ASCA observations
of the cluster of galaxies Abell 3627 are presented. 
This cluster, located behind the Milky Way, was recently found to be a nearby, 
X-ray bright and very rich cluster.
Pointed observations onto the central region of the cluster 
gave a gas temperature of $\sim7$ keV and a metallicity of about 0.2 solar.
An offset pointing to a substructure elongated to the south-east of the cluster center 
gave a significantly lower temperature of $\sim 5$ keV.
The 2--10 keV luminosity within a radius of $40'$ (1.1 Mpc) is 
estimated to be $3.7 \times 10^{44}$ erg s$^{-1}$.
The X-ray data imply a cluster mass of about $4\times 10^{14}$ $M_\odot$ within $40'$.

}

\kword{
Galaxies: abundances --- Galaxies: clustering ---
Galaxies: clusters: individual: (Abell 3627) --- X-rays: galaxies
}

\maketitle

\section{Introduction}
Abell 3627 (hereafter A3627) was recently found to be a nearby ($z=0.016$), very 
rich cluster behind the southern Milky Way by Kraan-Korteweg et al. (1996).
Although it was catalogued as a nearby rich cluster in the ACO catalog (Abell
et al. 1989), this cluster had received little attention before.
This is partly because of a high degree of galactic obscuration towards the 
cluster direction of ($l, b$) = (325$^{\circ}$,$-7^{\circ}$).
The galaxy velocity dispersion of $\sigma = 897 $ km s$^{-1}$ from a redshift 
survey (Kraan-Korteweg et al. 1996) indicates the gravitational mass of A3627 to be  
$5.1\times 10^{15}$ $M_\odot$, which is comparable to that of the Coma cluster, 
and is typical of a rich cluster.
The direction and distance of A3627 are also interesting, because the implied 
position falls near to the location predicted for the center of the Great 
Attractor (Lynden-Bell et al. 1988), suggesting that the cluster lies  
near the bottom of the Attractor's gravitational potential well.

This cluster had escaped X-ray detection in the previous surveys,
and was first discovered by the ROSAT All Sky Survey as one of the brightest 
clusters in X-rays.
A follow-up pointed observation with the ROSAT PSPC by B\"ohringer et al. 
(1996; hereafter B96) showed that the cluster is indeed very massive in both 
gas mass and gravitational mass.
The PSPC image shows that the cluster is elongated to the south-east direction.
The X-ray morphology of the cluster can be interpreted as a 
merger where a smaller cluster in the south-east part is colliding with the 
main cluster (B96). According to a PSPC observation (B96), 
the X-ray luminosity in 0.1--2.4 keV and the gas temperature are 
($2.2\pm 0.3)\times 10^{44} h_{50}^{-2}$ erg s$^{-1}$ and 5--10 keV, respectively, 
where the Hubble constant is expressed as $H_{\rm 0} = 50 h_{50}$ km s$^{-1}$ Mpc$^{-1}$.
However, the limited energy band of ROSAT and a large interstellar hydrogen column 
(around $1.8 \times 10^{21}$ cm$^{-2}$) prevented accurate determinations 
of the X-ray spectroscopic properties.

We observed the central regions of A3627 with ASCA (Tanaka et al. 1994) in the 
energy band of 0.5--10 keV.
In this article we report on the results of the properties of A3627 
obtained from these observations, and compare them with those of other 
nearby rich clusters of galaxies.

\section{Observations and Results}
\subsection{Observations}

The ASCA X-ray telescope (XRT; Serlemitsos et al. 1995) was pointed at the center 
of the cluster on 1996 February 27 and 28,
and $19'$ south-east from the center on 1996 February 29.
Hereafter, we refer to these two pointings as CENTER and SE.
The pointing directions were $16^{\rm h} 15^{\rm m} 44^{\rm s}, 
-60^{\circ } 49' 56''$ (J2000) and $16^{\rm h} 17^{\rm m}25^{\rm s}, -61^{\circ } 
4'13''$ (J2000) during the CENTER and SE observations, respectively.
The observations were performed with the SIS (Solid-state Imaging Spectrometer; 
Burke et al. 1994) in the 2--CCD faint/bright mode and the GIS (Gas Imaging 
Spectrometer; Ohashi et al. 1996) in the PH normal mode.
We have screened events by using standard data-selection criteria;
i.e., geomagnetic cutoff rigidity $>  6$ and $>$ 8 GeV {\it c}$^{-1}$
for the SIS and GIS, respectively, 
the elevation angle above the earth's horizon $> 10^\circ$ for the GIS, 
and $> 10^\circ$ (dark earth) or $> 25^\circ$ (sunlit earth) for the SIS\@. 
The thus-obtained net exposure times for the CENTER pointing are $\sim 29$ ks with 
the SIS and $\sim 37$ ks with the GIS,
while those for SE are $\sim 26$ ks with the SIS and $\sim 25$ ks with the GIS.

\subsection{The X-Ray Image}
The GIS count rates in 0.7--10 keV within the entire field of 
view were 2.5~ {c s$^{-1} / $ detector} for the CENTER pointing and
1.2~ {c s$^{-1}/$ detector} for the SE pointing.
These include background of 0.2~ {c s$^{-1}/$ detector}.
Figure 1 shows the background-subtracted mosaic X-ray image of A3627,
obtained with the GIS (GIS\,2$+$GIS\,3) from the two pointings.
The image was corrected for the exposure time, but not for the vignetting or
the supporting grids of the GIS window.
The X-ray brightness centroid is near to the radio galaxy PKS\,1610--60,
which is one of the three bright central galaxies of the cluster (B96),
 and has 
relatively large radio lobes.
The overall X-ray morphology is not circularly symmetric, as already recognized 
in the ROSAT image (B96).
The X-ray emission of the cluster almost fills the entire GIS fields of view 
of the two pointings and can be observed out to at least $40'$ (1.1 $h_{50}^{-1}$ 
Mpc ) toward the south-east direction from the center.
In the ROSAT observations, the cluster X-ray emission is found up to a radius of 
1$^{\circ}$ \ (B96).

\subsection{The X-Ray Spectra}
In order to study the spatially averaged properties of the X-ray emission,
we first utilized the data obtained with the GIS, 
 which cover a larger field of view compared to that of the SIS.
We accumulated the GIS events from the two detectors (GIS\,2 and GIS\,3) within a
circle having a projected radius of $14'$, one for each pointing, as shown in figure 2 as G-a and G-b.
The background spectrum was obtained from the ``blank-sky'' (containing no bright 
sources) database by extracting the events within an identical 
region for the same detector.
The thus-derived GIS spectra are shown in figure 3, 
without any correction for the 
energy-dependent instrument response. 
The circle for the CENTER region includes 
PKS\,1610--60. However, its contribution is estimated to be $\sim1$\% of the total 
photon number from the PSPC observation; hence, the effect to the observed spectrum 
is negligible.

We fitted the GIS spectra in 0.7--10.0 keV with 
an isothermal Raymond-Smith (1977) plasma model including the photoelectric 
absorption.
As shown in table 1 and illustrated in figure 3, the fits are generally acceptable.
The result shows that the temperature in the SE region is lower than that in the 
CENTER region by about 2 keV\@.
We find that the best-fit hydrogen-column densities are lower than the 
published galactic value of $1.8\times 10^{21}$ cm$^{-2}$ towards A3627 
obtained from the 21cm observations (Stark et al. 1992).
We return to this problem later.

Here, we should note that the obtained spectra contain the effect of the stray light,
i.e. photons scattered in from regions outside of the fields of view.
Although the stray light from the outer region to the CENTER region is negligible,
that from the central region to the SE region accounts for about 15\% of the 
observed flux.
In order to correct for this effect, 
we estimated the spectrum of the stray light from the central region to the SE region 
by a ray-tracing simulation (Honda et al. 1996)
employing the $\beta$ 
model brightness distribution derived from the ROSAT observations ($\beta = 0.56$ 
and a core radius of $r_{\rm c}=10'.0$; B96).
By subtracting the thus-estimated stray spectrum of the contamination,
we fitted the spectrum of the SE region again.
The temperature of the SE region turned out to be 4.0--5.3 keV\@.
This range is consistent with, but even lower than, that given in table 1.
Therefore, we conclude that the SE region has a lower temperature than the CENTER region,
even when taking into account the effects of stray light.

We next utilized the data from the SIS, which has a higher sensitivity at lower 
energies, hence allowing a more accurate determination of $N_{\rm H}$. 
We accumulated the SIS events within a circle of a projected radius $4'.9$,
one from CENTER and two from SE, as shown in figure 2.
We fitted the same model to the SIS spectra as that used for the GIS spectra.
As shown in table 1 and illustrated in figure 4, the fits are acceptable,
yielding the best-fit $N_{\rm H}$ values in the range $1.3-2.0\times 10^{21}$ cm$^{-2}$.
Within the statistical uncertainties, these are consistent with the above-mentioned 
galactic value, and also with the ROSAT PSPC result for which the 90\% limits span 
a range of $1.4 - 1.9\times 10^{21}$ cm$^{-2}$ (B96).
We consider that the lower values of $N_{\rm H}$ derived from the GIS spectra are probably 
due to a calibration uncertainty in the low-energy response of the GIS\@.
The temperature and metallicity derived from the SIS are fully consistent with 
those from the GIS, though with slightly larger errors.

The metallicity values determined from the iron K-line for different regions (table 1) are consistent with one another, 
showing a fairly uniform metallicity throughout the whole cluster.
To better determine the metallicity around the cluster center,
we fitted the spectra within a circle of radius of $5'$ from the cluster center
of the GIS and the SIS jointly with the isothermal plasma model.
As a result, the 90\% confidence limit of the metallicity was found to be 0.18--0.26 solar. 
We also obtained the 90\% confidence limit of the temperature from a joint 
fit to be 6.4--6.9 keV.

The 2--10 keV luminosity within a radius of $40'$ was estimated to be $3.7\times 
10^{44} h_{50}^{-2}$ erg s$^{-1}$ after correcting for absorption.
For this calculation we employed the spatial gas distribution derived from the ROSAT 
PSPC observations (B96) and the best-fit spectral model which we mentioned above.

\section{Discussion}

We derived the X-ray luminosity and the gas temperature in the central region 
of A3627 using the broad-band imaging spectroscopic data of ASCA.

A temperature of $6.7\pm 0.3$ keV, found in the CENTER region, is consistent 
with a velocity dispersion of 897 km s$^{-1}$ by Kraan-Korteweg et al. (1996).
The ratio of the specific energy in galaxies to that in the hot gas,
$\beta _{{\rm spec}} = 0.82$, is a typical value for clusters (Edge et al. 1991).
The X-ray luminosity versus temperature of A3627 is also in general agreement 
with the empirical correlation between these two quantities (David et al. 1993).
Assuming that the cluster has an isothermal and spherically symmetric gas 
distribution, the temperature measured in the CENTER region gives 
a gravitational mass of $4\times 10^{14} h_{50}^{-1} $ $M_\odot$ \ 
within $40'$ (1.1 $h_{50}^{-1}$Mpc) and a gas mass fraction 
(ratio of the gas mass to the total mass) of 11$h_{50}^{-1.5}$\%.
Here, we used again the gas-distribution model from the ROSAT observations (B96).
The gravitational mass of the main cluster extrapolated to 3 Mpc, which
is approximately the virial radius of this galaxy system, is about $10^{15}$
$M_\odot$ within an uncertainty of about a factor of 2. Thus, the main
component turns out to have about half the mass of the Coma cluster.
The substructure extending to the south-east direction may add
up to 50\% to the total cluster mass.
This total cluster mass is several times lower than that estimated from 
the galaxy velocity dispersion by Kraan-Korteweg et al. (1996). However, this 
is not a discrepancy, but is simply because their mass refers to a much larger 
volume.

We found that the temperature of the SE region, $4.0-5.3$ keV, is significantly 
lower than that of the CENTER region, which confirms the temperature drop in 
the south-east direction from the cluster center previously noted by B96. 
The temperature difference suggests that there is 
a systematic temperature drop toward the outer region, 
or that the SE region is a substructure that is 
cooler than the main cluster. 
The agreement of the average 
temperature within $14'$ radius with that for the central $5'$ radius 
supports a uniform temperature distribution in the main cluster 
within the observed scale. Therefore, the lower temperature seems to be a 
particular property of the SE region, which is coincident with the substructure 
in the X-ray brightness distribution (B96). Based on the morphology, 
B\"{o}hringer et al. (B96) suggested that this substructure is a subcluster 
in the process of merging with the main cluster. 

Together with the imaging data (B96), we may attempt to interpret the nature
of the merger configuration. 
The projected distance of the two merging subunits is about (0.5 -- 0.6) $h_{50}^{-1}$ Mpc. 
If the inclination angle between the sky plane and the merger axis is not very large, 
the two merging units are already in close contact. 
In principle, there are three alternative possibilities: 
(1) the merging subunit is on its first infall and not yet affected by the resulting shock;
(2) the infalling subunit is already subject to shock heating; or
(3) the two untis have already passed each other\@. 
In the last case, after the core collision, there is a rebounce phase which can lead to a separation of the two merging units on the order of 1 $h_{50}^{-1}$ Mpc at turnaround
(e.g. Burns et al. 1994).

In the first case, the two units are expected to behave as two independent clusters, 
following individually the normal luminosity ($L_{\rm X}$) versus temperature ($T$) scaling of 
$L_{\rm X} \propto T^3$ (e.g., David et al. 1993)\@.
However, the observed flux ratio between the main and sub components, 
approximately 7 -- 10 : 1, 
is much larger than the ratio of $\sim 3$ predicted by their temperature ratios.
Therefore, this possibility may be unlikely.
However, in the third case
the subcomponent has already passed through the main cluster center, 
and the subcomponent gas was heated and reassembled in the subcluster potential 
(see e.g. Schindler, M\"uller 1993; Kulp 1998).
In this case, it is not surprising that the two temperatures are comparable. 
However, the merging subunit in the south-east still appears to be fairly compact, and the central  
radio-lobe galaxy does not exhibit much disturbance. 
Therefore, this alternative is not very likely, either.
Thus, by elimination, 
the second possibility is the most plausible.

Honda et al. (1996) also found a significant temperature variation in the Coma Cluster. 
It is interesting to note that its prominent 
subcluster in the south-west direction, centered on NGC 4839, also shows a lower temperature than that of the cluster average. 
Another low-temperature region of the Coma cluster in the south-east actually 
coincides with a substructure associated with a bright elliptical NGC 4911.
This is one of the clear substructural features noted by Briel et al. (1992)
and later by others (White et al. 1993; Vikhilinin et al. 1997). 
These facts appear to be similar to the case of A3627.

The ASCA observations enabled us to determine the metallicity of the 
ICM of A3627 for the first time.
The spatially averaged metallicity of 0.16--0.31 solar agrees with the 
typical values of 0.15--0.3 solar in rich clusters,
including the Coma cluster, observed with Ginga (Hatsukade 1989) and with ASCA (e.g., Fukazawa et al. 1997).
The results from the CENTER and SE regions are consistent with a uniform abundance 
distribution over the entire cluster.   
We found no significant metallicity increase toward the center of the 
cluster, which had been found in some low-temperature clusters,
particularly those 
containing cD galaxies (Koyama et al. 1991; Fukazawa et al. 1994; Xu et al. 
1997; Ezawa et al. 1998).

It is quite interesting that we found this very massive cluster at a distance of less than 
100 $h_{50}^{-1}$ Mpc, approximately in the direction
of the Great-Attractor. 
Even though the cluster fails by more than an 
order of magnitude to provide the mass for the Great-Attractor effect,
we may recall that similarly massive nearby clusters, 
Coma and Perseus, are found in the Great Wall superstructure and the Perseus-Pisces
supercluster, respectively; 
two structures with estimated total masses on
the order of $10^{16}$ $M_{\odot}$. Therefore, the discovery of this massive
cluster and its confirmation by X-ray observations may indeed suggest a way
of finding additional massive structures contributing to the Great Attractor 
still hidden by the Milky Way.

\vspace{1.5cm}
We thank D. M. Neumann, S. Schindler, and M. Hirayama for helpful discussions 
and the ASCA-ANL and SimASCA software development teams and members of the 
ASCA team for making this study possible.

\begin{table}[p]
\begin{center}
\caption{Results of the single-temperature Raymond-Smith model fits.$^*$}
\label{tbl:sp_fit}
\begin{tabular}{lccccccccc}
\hline
\hline
$\natural$& Pointing	& Energy	& Sensor & $R^\sharp$	& $N_{\rm H}$	& $kT$    	
& $A^\dagger$		& $\chi ^2/\nu$ \\
		  & 		& (keV)		& 	 & 		& 
$(10^{21}$cm$^{-2}$)& (keV)	&			& \\	
\hline
G-a	& CENTER	& 0.7--10.0	& G2+G3		& 14	& 1.0 (0.9--1.2)	& 6.7 
(6.4--7.0)	& 0.16 (0.13--0.20)	& 134.6/104\\
S-a	& CENTER	& 0.6--10.0	& S0C1/S1C3	& 4.9	& 2.0 (1.9--2.1)	& 7.1 
(6.8--7.5)	& 0.23 (0.17--0.29)		& 343.4/331\\
G-b	& SE		& 0.7--10.0	& G2+G3		& 14	& 0.7 (0.3--0.8)	
& 5.1 (4.9--5.6)	& 0.25 (0.20--0.31)	& 111.0/104\\
S-b	& SE		& 0.6--10.0	& S0C2/S1C0	& 4.9	& 1.9 (1.6--2.1)	
& 5.9 (5.4--6.7)	& 0.31 (0.20--0.45)	& 282.6/206\\
S-c	& SE		& 0.6--10.0	& S0C1/S1C3	& 4.9	& 1.3 (1.0--1.5)	
& 5.2 (4.6--5.7)	& 0.27 (0.14--0.42)	& 202/179\\
\hline
\end{tabular}
\end{center}
\begin{description}
\begin{footnotesize}
\setlength{\itemsep}{-1mm}
\item[$^*$] The error regions in parentheses refer to single-parameter 
           90\% confidence limits. 
\item[$^\natural$] The region name in figure 2. The center of regions are
G-a) $16^{\rm h} 14^{\rm m}57^{\rm s}, -60^{\circ } 51'35''$,
G-b) $16^{\rm h} 17^{\rm m} 28^{\rm s}, -61^{\circ }4'10''$,
S-a) $16^{\rm h} 14^{\rm m}53^{\rm s}, -60^{\circ } 55'22''$,
S-b) $16^{\rm h} 16^{\rm m}50^{\rm s}, -60^{\circ } 58'20''$, and
S-c) $16^{\rm h} 16^{\rm m}34^{\rm s}, -61^{\circ } 9'36''$.
\item[$^\sharp$] The integrated radius in arcmin.
\item[$^\dagger$] Overall metallicity in solar unit mainly determined by the Fe-K line. 
Solar abundance ratios are assumed. The solar Fe/H ratio is taken to be $4.68 \times 10^{-5}$ by 
number.

\end{footnotesize}
\end{description}
\end{table}

\newpage
\section*{References}
\re
Abell G. O., Corwin H. G., Olowin R. P.\ 1989, ApJS 70, 1
\re
B\"ohringer H., Neumann D. M., Schindler S., Kraan-Korteweg R. C.\ 1996, ApJ 467, 168
\re
Briel U. G., Henry J.P., B\"ohringer H.\ 1992, A\&A 259, L31
\re
Burke B. E., Mountain R. W., Danies P. J., Dolat V.S.\ 1994, IEEE Trans. Nuc. Sci. 41, 375
\re
Burns J. O., Roettiger K., Ledlow M., Klypin A.\ 1994, ApJ 427, L87
\re
David L. P., Slyz A., Jones C., Forman W., Vrtilek D., Arnaud K.\ 1993, ApJ 412, 479
\re
Edge A. C., Stewart G. C.\ 1991, MNRAS 252, 428
\re
Ezawa H., Fukazawa Y., Makishima K., Ohashi T., Takahara F., Xu H., Yamasaki N. Y.\ 1998, ApJ 490, L33
\re
Fukazawa Y., Ohashi T., Fabian A. C., Canizares C. R., Ikebe Y., 
      Makishima K., Mushotzky R.F., Yamashita K.\ 1994, PASJ 46, L55
\re
Fukazawa Y., Makishima K., Tamura T., Ezawa H., Xu H.,  Ikebe Y., 
      Kikuchi K., Ohashi T.\ 1997, PSAJ 50, 187
\re
Hatsukade I.\ 1989, PhD Thesis, Osaka University
\re
Honda H., Hirayama M., Watanabe M., Kunieda H., Tawara Y., Yamashita K., Ohashi T., Hughes J. P., Henry J. P.\ 1996, ApJ 473, L71
\re
Koyama K., Takano S., Tawara Y.\ 1991, Nature 350, 135
\re
Kraan-Korteweg R. C., Woudt P. A., Cayatte V., Fairall A. P., Balkowski C., Henning P. A.\ 1996, Nature  379, 8
\re
Kulp K. 1998, Diploma Thesis, Universit\"at M\"unchen 
\re
Lynden-Bell D., Faber S. M., Burstein D., Davies R. L., Dressler A., Terlevich R. J., Wegner G.\ 1988, ApJ 326, 19
\re
Ohashi T., Ebisawa K., Fukazawa Y., Hiyoshi K., Horii M., Ikebe Y., Ikeda H., Inoue H. et al.\ 1996, PASJ 48, 157
\re
Raymond J.C., Smith B.W.\ 1977, ApJS 35, 419
\re
Schindler S., M\"uller E. 1993, A\&A 272, 137
\re
Serlemitsos P. J., Jalota L., Soong Y., Kunieda H., Tawara Y., Tsusaka Y., Suzuki H.,
	Sakima T. et al.\ 1995, PASJ 47, 105
\re
Stark A.A., Gammie C.F., Wilson R.W., Bally J., Linke R.A., Heiles C., Hurwitz M.\ 1992, ApJS, 79, 77 
\re
Tanaka Y., Inoue H., Holt S. S.\ 1994, PASJ 46, L37
\re
Xu H., Ezawa H., Fukazawa Y., Kikuchi K., Makishima K., Ohashi T., Tamura T.\ 1997, PASJ 49, 9
\re
Vikhlinin A., Forman W., Jones C.\ 1997, ApJ 474, L7
\re
White S. D. M., Briel U. G., Henry J. P.\ 1993, MNRAS 261, L8

\newpage
\noindent
Fig.1.\\
X-ray image of Abell 3627 taken with the GIS (GIS\,2+GIS\,3) in
0.7--10 keV, synthesized from two partially overlapping pointings.
The image was smoothed using a Gaussian filter with $\sigma = 0'.5$, but not
corrected for the XRT vignetting or partial shadows due to the detector support
ribs. 
The contour levels are logarithmically spaced from $10^{-5}$ c s$^{-1}/$ pixel to $10^{-3}$ c s$^{-1}/$ pixel.
The sky coordinates are J2000.
\vspace{1cm}

\noindent
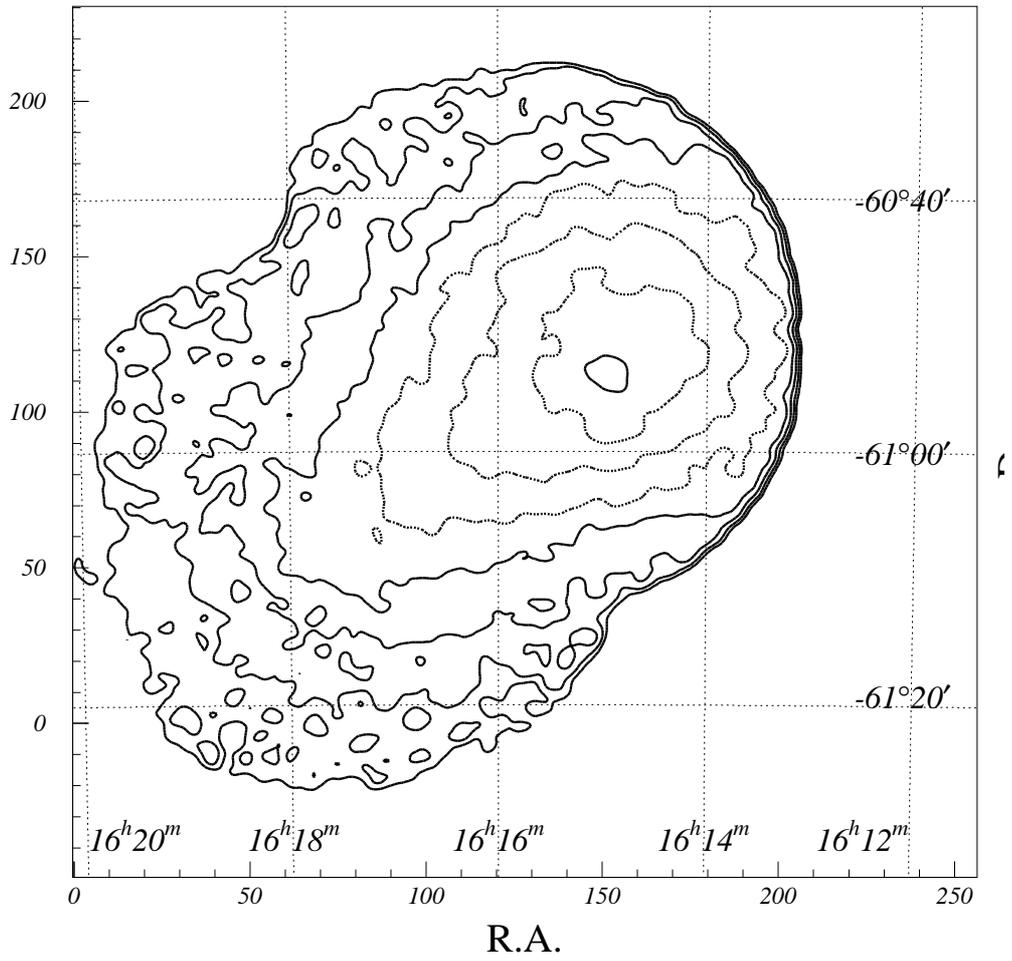
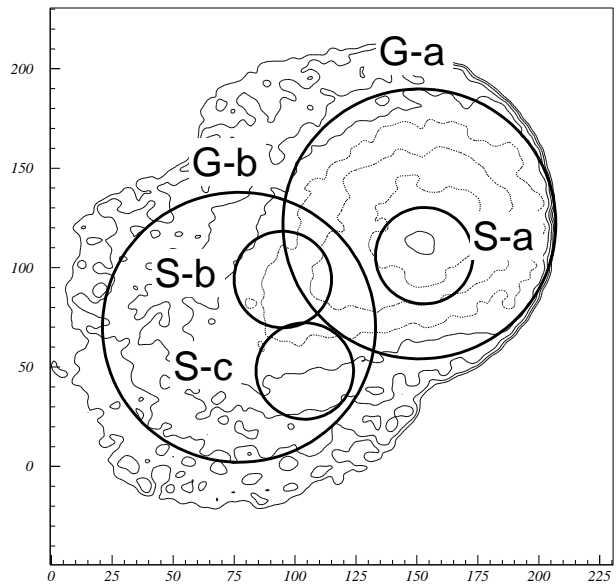
Fig.2.\\
Five regions from which the energy spectra were accumulated.
\vspace{1cm}

\noindent
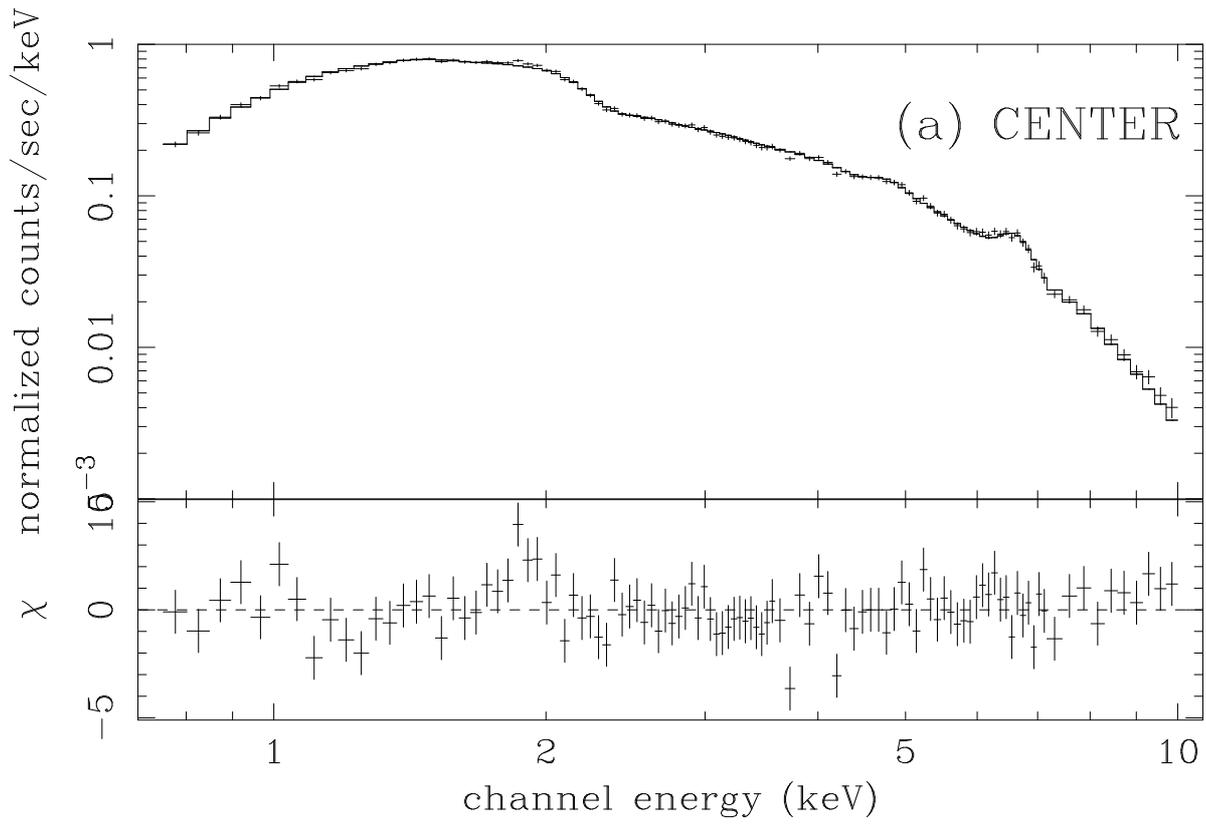
Fig. 3.\\
GIS spectrum (sum of the two detectors) of A3627 in (a) CENTER region and (b) SE region,
shown together with the best-fit Raymond-Smith model 
and the fit residuals (see table 1).

%
\vspace{1cm}

\noindent
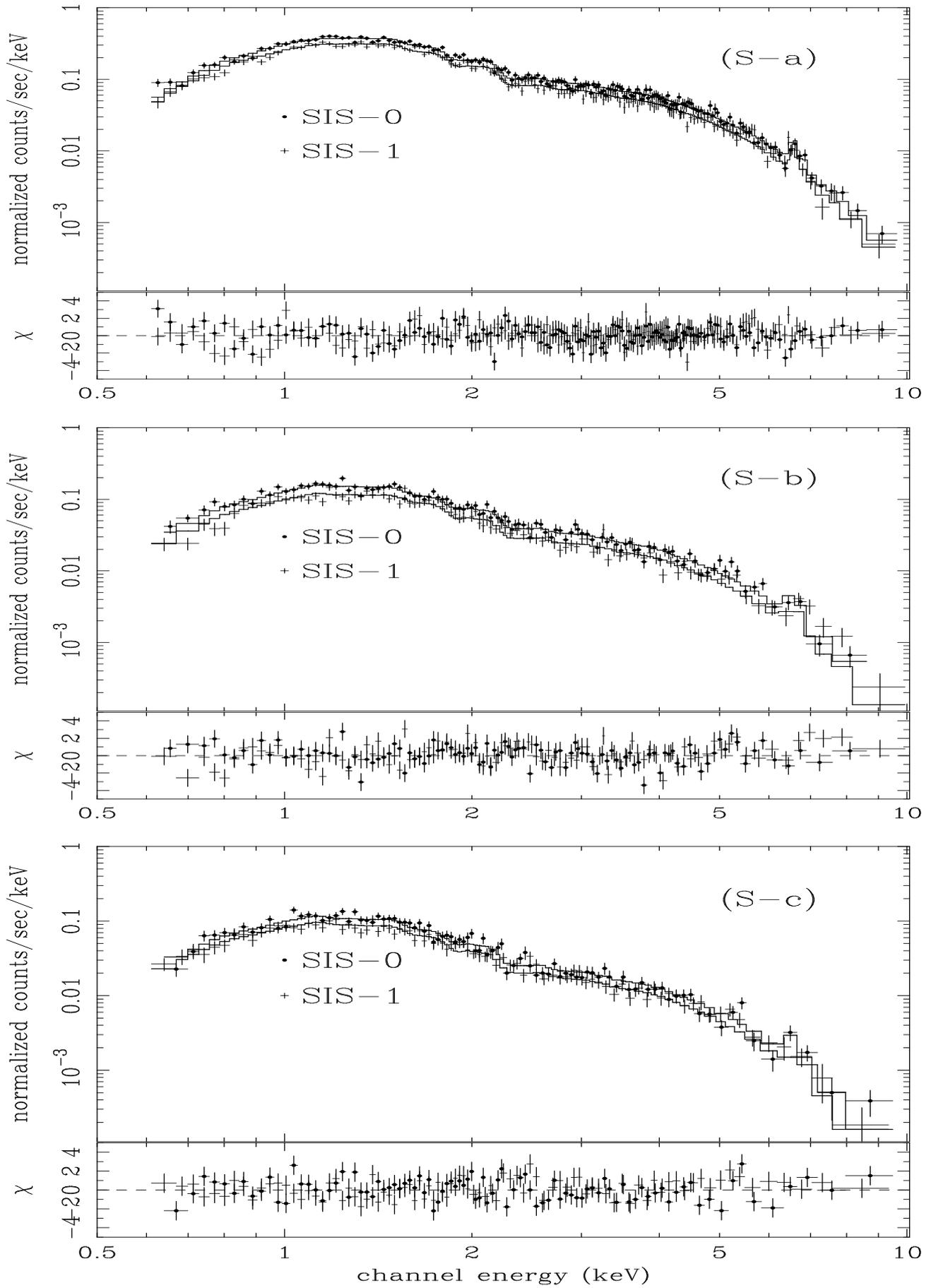
Fig. 4.\\
SIS spectrum of A3627 in each region of figure 1,
shown together with the best-fit Raymond-Smith model 
and the fit residuals (see table 1).
The spectra from SIS-0 and SIS-1 are seprately shown.

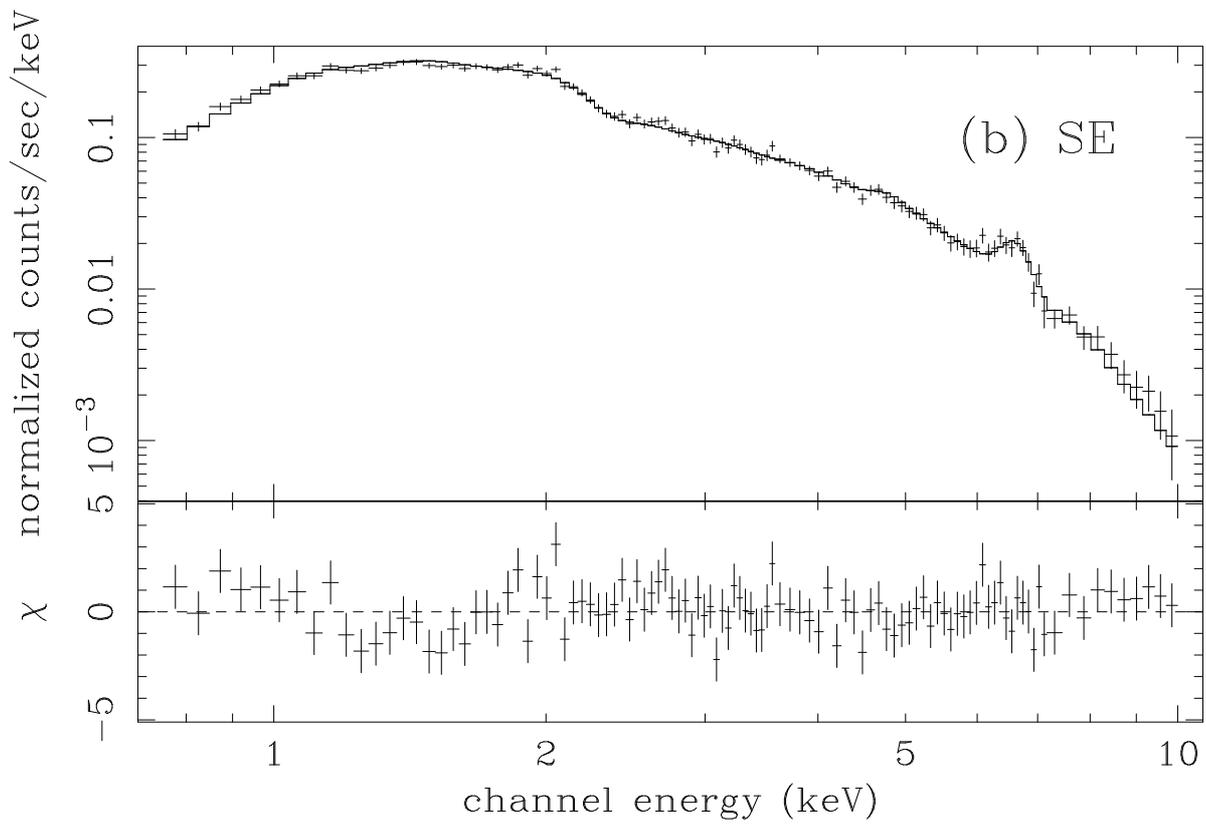
\end{document}